\documentclass[a4paper, 10pt]{article}
\usepackage[margin=2.5cm]{geometry}

\usepackage{graphicx, cite, booktabs, url}
\graphicspath{ {./_figures} }

\PassOptionsToPackage{hyphens}{url}\usepackage{hyperref}
\hypersetup{colorlinks=true, urlcolor=blue, citecolor=blue}
\usepackage[labelfont=bf,labelsep=period,justification=justified, singlelinecheck=false]{caption}
\usepackage{amssymb, amsmath}
\usepackage{setspace}
\usepackage{placeins} 
\title{\textbf{Trends in Equal-Contribution Authorship: A Large-Scale Bibliometric Analysis of Biomedical Literature}}
\author{Binbin Xu}
\date{\normalsize EuroMov Digital Health in Motion, Univ. Montpellier, IMT Mines Ales, France\\
\texttt{binbin.xu@mines-ales.fr}}

\begin{document}

\maketitle

\begin{abstract}
Equal-contribution authorship, in which two or more authors are designated as having contributed equally, is increasingly common in scientific publishing. Using approximately 480,000 tagged records from PubMed and PMC (2010-2024), we examine temporal trends, journal-level patterns, geographic distributions, and byline positions of equal-contributing authors. Results show a sharp rise after 2017, with both high-output mega-journals and smaller, discipline-specific journals contributing to the growth. Journal-level analysis indicates a median increase in the share of tagged articles from about 19\% in 2015 to over 30\% in 2024, with some journals exceeding 50\%. Geographically, China accounts for the largest share (40.8\% of fractionalized contributions), followed by the United States (15.2\%) and Germany (5.2\%). Normalizing to 2015 baselines, China shows a $13.1\times$; increase by 2024, while even the slowest-growing countries more than tripled their levels. Analysis of normalized byline positions shows that equal-contribution designations are concentrated near the first-author position, with fewer cases in middle or last positions. These findings document a broad shift toward shared first-author credit across journal sizes and regions within the biomedical literature and suggest that journals and evaluators may need to rely more on transparent contributorship information and to monitor the use of such labels over time.
\end{abstract}

\begin{flushleft}
\end{flushleft}

\section{Introduction}

Over the last two decades, scientific publications have increased at an almost exponential rate. A recent study showed that the total number of articles indexed in Scopus and Web of Science was close to 2.9 million in 2022, ~47\% higher than in 2016 and almost doubled the figure in 2013 \cite{Hanson2024strain}. However, this represents only a small portion of the total scientific publication. As of 2025, PubMed alone indexes over 38 million references. 
Not only has the number of publications grown rapidly, but there has also been a clear shift from solo scholarship toward team-based research. In biomedical fields, this growth in team size has been accompanied by new conventions to highlight contributions and responsibilities, including explicit designation of \texttt{"equal contribution"} (often in co-first authorship) and, more recently, co-corresponding authorship. Early journal-specific studies documented a clear increase in equal-contribution statements in high-impact medical journals between 2000 and 2009 and in specialty venues thereafter \cite{Akhabue2010Equal}.

Teams have expanded substantially across science: typical team size roughly doubled over the latter half of the twentieth century and continues to grow \cite{Wuchty2007Increasing,Fortunato2018Science}. In biomedicine, authors per article increased from 3.99 (2000) to 6.25 (2020), while single-author papers declined about threefold \cite{Jakab2024How}. Growth includes ``hyperauthorship'' (very large bylines) with attendant attribution challenges \cite{Cronin2001Hyperauthorship}. Collaboration correlates with higher citation impact but exhibits diminishing marginal returns at very large team sizes \cite{Lariviere2015Team}. The article-level citations also tend to rise with author and reference counts in ecology research, consistent with incentive effects without implying causation \cite{Fox2016Citations}. Within biomedicine, the division of labor has produced an expanding stratum of ``middle authors'' \cite{Mongeon2017rise}. These macro-trends provide neutral context for the increased use of explicit equal-contribution designations examined here.

A large body of work has examined how credit and responsibility are allocated in multi-author papers. Classic editorials and surveys distinguished guest, gift, and ghost authorship and documented their presence in biomedical journals, raising concerns about fairness and accountability in author lists \cite{Rennie1994Authorship,Flanagin1998Prevalence}. Later reviews and empirical studies suggest that inappropriate authorship practices remain common despite formal guidelines and that junior researchers may be particularly exposed to coercive or honorary authorship arrangements \cite{Kornhaber2015Ongoing,PardalRefoyo2025Gift,Khalifa2022Losing}. Qualitative and conceptual analyses further highlight the role of local norms and power relations in authorship decisions, including the strategic control of prestigious first and last positions and the so-called ``White Bull'' effect, in which senior researchers systematically accumulate credit from collaborators \cite{Kwok2005White,Macfarlane2015ethics}.

Within this more general context, equal-contribution and joint first authorship have received specific attention. Interview-based studies of authors who share first authorship report both perceived benefits, such as recognition of genuinely joint leadership, and tensions arising from ambiguous interpretations of these labels by evaluation committees \cite{Hosseini2020Qualitative}. Recent commentaries and editorial discussions ask whether joint or co-first authors are treated as truly equal in practice and how such designations interact with CV presentation and promotion criteria \cite{Owens2024Equality,Efron2024Joint}. More generally, work in the sociology and history of science argues that emerging authorship formats, including contributorship taxonomies and new forms of credit attribution, are part of a more general credit / responsibility reconfiguration in science \cite{Biagioli2012Recycling,Biagioli2022Ghosts}. The present study contributes to this line of research by providing a large-scale, metadata-based characterization of equal-contribution authorship across journals and countries, without making assumptions about the appropriateness of individual cases.

These authorship signals have prompted the development of guidance from editorial bodies and standards organizations. 
For example, the International Committee of Medical Journal Editors (ICMJE) formalized authorship criteria and encouraged contributorship statements to improve transparency around individual roles \cite{ICMJE2025Defining, ICMJE2025Recommendations}. The CRediT taxonomy \cite{NISOCRTWG2022ANSINISO} introduced a structured vocabulary of 14 roles for describing individual contributions, which has now been adopted by many publishers. 
Major publisher policies now define how \texttt{equal contribution} is recorded. In practice, many publishers such as Nature, BMJ, PLOS align with ICMJE recommendations on authorship and require contributorship statements using the CRediT taxonomy so that equal-contribution designations are explicitly declared rather than inferred \cite{ NaturePortfolio2025Authorship, BMJ2025Authorship, PLOS2025Authorship}. Policies typically cap formal equal-contribution labels (e.g., a single co-first and a single co-senior set) and direct additional equalities to the contribution statement, though implementation varies by journals. 

In this study we treat such labels as descriptive metadata (not evaluative) and harmonize cross-publisher heterogeneity via JATS/PubMed tags. At the metadata level, the Journal Article Tag Suite \cite{Beck2011NISO} (JATS) and PubMed/PMC pipelines \cite{Central2025PubMed} support machine-readable flag, e.g., the \texttt{equal-contrib="yes"} attribute on \texttt{<contrib>} and explicit corresponding-author elements. They provide a technical basis for corpus-scale measurement. 

Despite these developments, the literature lacks a comprehensive, longitudinal account of equal-contribution authorship across the biomedical record. Existing studies typically focus on selected journals, a single discipline, or short time windows, and often rely on manual searches or unstructured PDF text. 
For example, Akhabue and Lautenbach \cite{Akhabue2010Equal} quantified equal-credit statements in five general medical journals (2000-2009), and subsequent analyses reported similar growth in selected specialties. Huang et al. \cite{Huang2016co} examined 53,603 pharmacy and anesthesia papers (1995-2014) from top journals to quantify the prevalence, distribution, and byline patterns of co-first and co-corresponding authorship. Related work on co-corresponding authorship \cite{Tian2024Understanding}, a different yet related designation of responsibility, confirms that multi-role attribution is spreading across countries, disciplines, and journals, but comprehensive coupling analysis using large-scale structured metadata remains lacking. 

This gap has implications for research assessment. Committees, project funders, and hiring panels often rely on author order as a proxy for leadership and contribution. As shared first authorship becomes more common, the interpretive value of ordered position may diminish unless explicitly supplemented by author statements. 

Here, we present a corpus-level analysis of equal-contribution authorship from 2010-2024 using machine-readable metadata from PubMed and PubMed Central. Leveraging JATS/PubMed tags (e.g., values of \texttt{equal-contrib} on \texttt{<contrib>}), we (i) quantify global trends in the prevalence of papers with $\geq2$ equal-contribution authors; (ii) map journal- and year-specific patterns; and (iii) examine geographic patterns via affiliation data. This approach complements prior journal- or domain-focused studies by exploiting standardized metadata across a large collection of biomedical literature. By establishing a longitudinal baseline, this work documents the current state and recent evolution of equal-contribution authorship and provides a reference point for future studies on co-corresponding authorship and other contributor-role designations.

The remainder of this article is organized as follows: Section~\ref{sec:data} describes the data sources and processing steps. Section~\ref{sec:results} presents the results, including overall trends, journal-level patterns, and geographic analysis. Section~\ref{sec:discussion} discusses the implications of these findings and concludes the article.

\section{Data}
\label{sec:data}

In the biomedical field, two major bibliographic databases exist: PubMed and PubMed Central (PMC), which are complementary services of the U.S. National Library of Medicine (NLM). PubMed is the largest free bibliographic database today indexing over 38 million citations and abstracts from MEDLINE, life-science journals, and online books. It also provides links to full text when available (including PMC) and offers annual XML baselines and updates\footnote{accessible on \url{https://ftp.ncbi.nlm.nih.gov/pubmed}}. 
PMC, on the other hand, contains a free full-text archive of biomedical and life-sciences journal literature, curated under NLM collection policies and distributed via FTP or cloud access (e.g., the Open Access Subset \cite{BNLM2025PMC}). Importantly for this study, PMC and PubMed support machine-readable contributor metadata in JATS/XML (e.g., the \texttt{equal-contrib="yes"} attribute on \texttt{<contrib>} and \texttt{corresp="yes"} for corresponding authors), enabling corpus-level measurement of equal-contribution authorship.

It is important to note that the two platforms are not systematically cross-linked: many PubMed records have no PMC counterpart and vice versa, and identifiers (PMID, PMCID, DOI) are sometimes missing or non-bijective. Accordingly, we match on available identifiers, de-duplicate overlaps, and exclude PMC \texttt{<sub-article>} content to curate a single, analysis-ready dataset.

\begin{figure}[!htb]
    \centering
    \includegraphics[width=\linewidth]{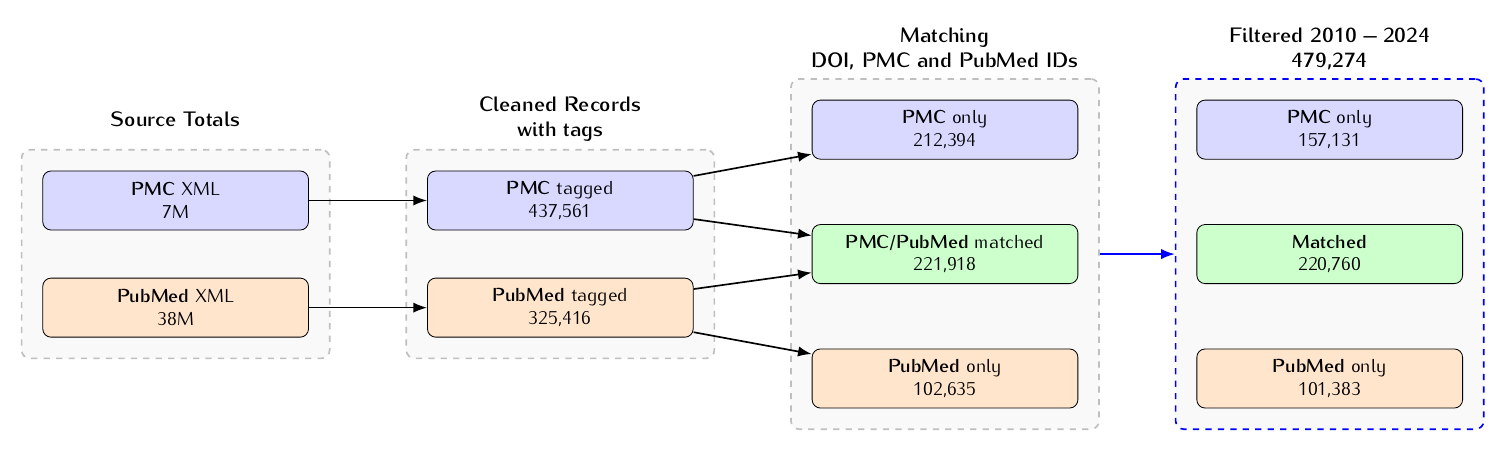}
    \caption{Data-flow diagram showing the filtering, matching, and date-restriction steps applied to PubMed and PubMed Central (PMC) records to produce the final analysis set for 2010-2024.}
    \label{fig:data_flow}
\end{figure}

A schematic of this workflow, including counts at each step and the identifier matching strategy, is shown in Figure \ref{fig:data_flow}. We processed PubMed and PubMed Central (PMC) in four stages. 
We began with the full corpora (PMC $\approx$ 7 million references, 146 GB compressed \texttt{.tar.gz} XML files; PubMed $\approx$ 38 million references, 46 GB compressed \texttt{.tar.gz} XML files), accessed in July 2025. 
The downloaded XML files containing contributor tags (e.g., equal-contribution indicators) were then extracted. Basic cleaning was applied, including discarding malformed XML, excluding \texttt{<sub-article>} translations/supplements, and removing obvious duplicates. This step yielded 437,561 PMC-tagged and 325,416 PubMed-tagged records.
In the third step, we reconciled overlaps into three subsets: PMC-only (212,394), PMC-PubMed matched (221,918), and PubMed-only (102,635) using DOI, PMCID, and PMID. 
Considering that earlier-stage records may not be well indexed with corresponding tags, we restricted the analysis window to 2010-2024. The final dataset contains 479,274 records in total, comprising 157,131 PMC-only, 220,760 PMC-PubMed matched, and 101,383 PubMed-only records for the longitudinal analyses.

\subsubsection*{Annual Publication Counts per Journal}
To normalize the yearly proportion of equal-contribution articles by journal, we computed the annual total number of publications for each journal in the combined dataset. This step required matching and merging bibliographic metadata from the raw PubMed and PMC collections while removing duplicates across sources. The PubMed dataset includes, for each article, fields such as journal abbreviation, publication date, and keys such as PMID, DOI, and PMC ID (if available). The PMC dataset contains similar bibliographic information, with multiple publication date fields (\texttt{pub\_date\_ppub}, \texttt{pub\_date\_epub}, \texttt{pub\_date\_pmc-release}). To avoid double counting, we applied a hierarchical deduplication strategy using DOI as the primary key, followed by PMID, and then PMC ID when DOI was unavailable. In cases of key collisions, PubMed entries were given precedence as the baseline source. For PMC records, the article date was defined as the earliest of the three available publication dates. For each record, we retained the journal identifiers (PubMed's \texttt{journal\_abbr} and PMC's \texttt{journal\_nlm\_ta}), the standardized article date, and the key bibliographic identifiers. This merged and deduplicated table was then used to generate annual publication counts per journal, which formed the denominator for calculating normalized yearly ratios of equal-contribution authorship.

\subsubsection*{Publication Country or Region Extraction}

We noticed that publications from certain regions tend to involve more team-based research, prompting investigation into whether geographic patterns exist in the prevalence of equal-contribution authorship. To examine this, we mapped the country or region of each author's affiliation.
Affiliation data were extracted separately for PubMed and PMC using custom parsing scripts tailored to their respective XML structures. For PubMed, the script iterated through \texttt{<AffiliationInfo>} elements, retaining the full affiliation string and isolating the last segment (typically the country or region). For PMC, affiliations were mapped to authors via \texttt{<xref ref-type="aff">} links, with a fallback to inline \texttt{<aff>} tags when no cross-reference was present. Text cleaning steps included removing label and email elements and standardizing whitespace. Country information was first taken directly from \texttt{<country>} tags when available; otherwise, it was heuristically inferred from the last segment of the affiliation string. 

In a post-processing step, extracted country names were matched against the ISO 3166-1 standard. Strict matches were assigned the corresponding ISO 3166-1 alpha-3 code. Remaining unmapped entries were manually reviewed, applying location-based heuristics such as city name lists and associated postal code patterns (notably for the USA, UK, and China). Affiliations that could not be resolved to a country, about 0.74\% of the total were left unassigned. Original affiliation strings were retained in all cases for traceability.

\subsubsection*{Data availability}

The final data has been deposited in Zenodo as closed-access records \cite{xu_2025_16968197}.

\section{Results}
\label{sec:results}

Please note that, all statistics reported here are descriptive; we do not infer policy intent, quality, or appropriateness.

From the total 479,274 records, as shown in Table~\ref{tab:eqc_corresp_distribution}, we first observed that the majority of {equal-contribution} designations involve two authors (73.4\%), with smaller fractions for three (17.0\%) and four authors (6.1\%). More extreme cases with five or more authors sharing first or equal-contribution authorship are rare but present, accounting for about 3.5\% of cases. The pattern for corresponding authors is similarly skewed: about two-thirds of articles list a single corresponding author, about 26\% list two, approximately 7\% list three or more. This skew toward smaller numbers suggests that while collaborative contributions are often recognized at the \emph{first author} level, the role of corresponding author remains more centralized, potentially reflecting editorial practices or the need for a single primary contact.

\begin{table}[!htb]
  \centering
  \caption{Distribution of the number of \texttt{equal-contribution} authors and corresponding authors across the combined PubMed-PMC dataset, 2010-2024.}
  \small
    \begin{tabular}{ccc|ccc}
    \toprule
    \multicolumn{3}{c|}{\textbf{Equal Contribution}} & \multicolumn{3}{c}{\textbf{Corresponding}} \\
    \midrule
    \textbf{Number} & \textbf{Count} & \textbf{Percent} & \textbf{Number} & \textbf{Count} & \textbf{Percent} \\
    \midrule
    2     & 351955 & 73.44\% & 1     & 316182 & 65.97\% \\
    3     & 81403 & 16.98\% & 2     & 128414 & 26.79\% \\
    4     & 29057 & 6.06\% & 3     & 29703 & 6.20\% \\
    5     & 8510  & 1.78\% & 4     & 4109  & 0.86\% \\
    6     & 3838  & 0.80\% & 5     & 620   & 0.13\% \\
    $\geq 7$ & 4511  & 0.94\% & $\geq 6$ & 246   & 0.05\% \\
    \bottomrule
    \end{tabular}%
  \label{tab:eqc_corresp_distribution}%
\end{table}%

\subsection{Trends in Equal-Contribution Authorship}

\begin{figure}[!htb]
    \centering
    \includegraphics[width=\linewidth]{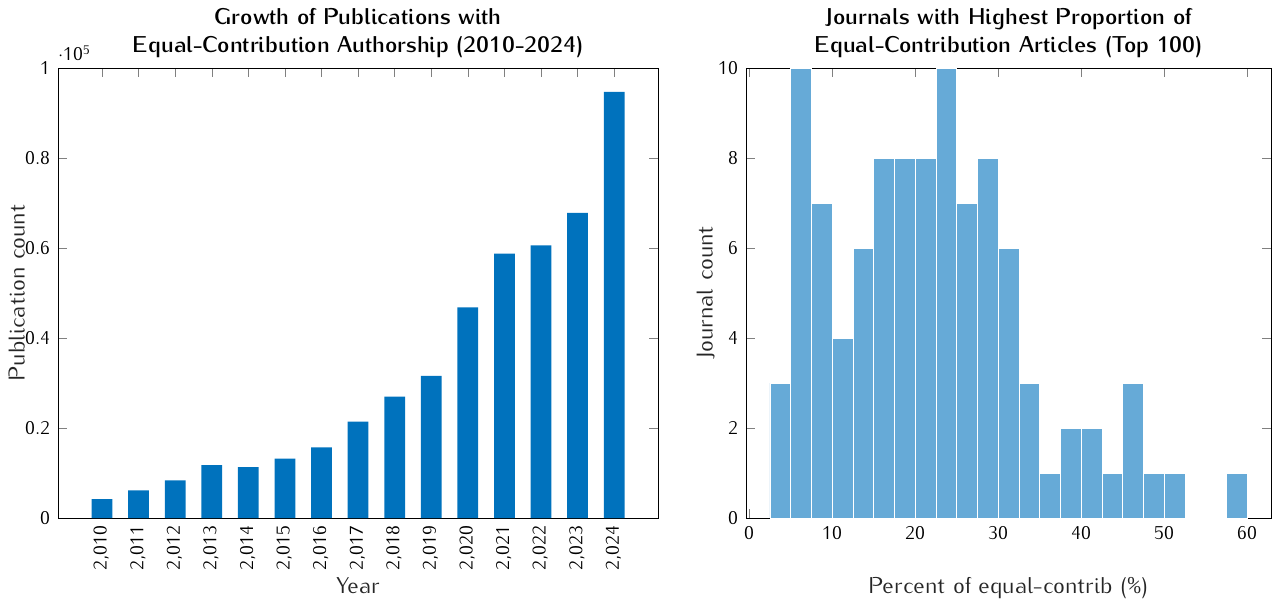}
    \caption{Trends in the use of equal-contribution authorship in scientific publications.\\
\emph{Left}: Annual count of publications explicitly labeling at least two authors as having contributed equally, showing a strong upward trend from 2010 to 2024.
\emph{Right}: Distribution of the top 100 journals ranked by proportion of equal-contribution articles (2010-2024).}
    \label{fig:journal_pub_year_counts_top}
\end{figure}

Figure \ref{fig:journal_pub_year_counts_top} shows a clear and accelerating increase in the absolute number of publications tagged with \texttt{equal-contribution} authorship between 2010 and 2024. In the left panel, the yearly totals rise gradually through the early 2010s, followed by a sharper increase after 2017. In the period after 2020, yearly totals peaked at $94,690$ tagged articles in 2024. This inflection would coincide with a broader uptake of explicit \texttt{equal-contribution} statements in journal metadata, which may reflect genuine growth in multi-author collaborative research, stronger reporting norms for shared first authorship, evolving journal policies, or changes in authorship arrangements.

The right panel of Figure \ref{fig:journal_pub_year_counts_top} focuses on journals in which \texttt{equal-contribution} authorship accounts for at least 15\% of all publications from 2010-2024. By normalizing for overall publishing volume during this period, this view highlights smaller but highly specialized journals alongside multidisciplinary mega-journals. While broad-scope journals typically fall in the 15-30\% range, several lower-volume journals reach much higher normalized ratios, with some exceeding 40-50\%. These high ratios suggest that in some domains, such as oncology, molecular biology, and genomics, \texttt{equal-contribution} authorship has almost become a common convention, potentially reflecting the multi-expertise, multi-center study designs and other collaborative research structures. Many of these high-ratio journals also contribute disproportionately to the steep post-2018 rise shown in the left panel, indicating that the overall trend is driven by both growth in mega-journals and increases within specialized journals.

Taken together, these two views highlight complementary aspects of the phenomenon: mega-journals contribute most to the total volume of equal-contribution articles, whereas smaller or more specialized journals can have a higher proportion of such designations. This dual pattern would suggest a broad evolution in authorship practices influenced by both journal scope and by research culture. Increasingly, modern studies are technically and logistically more complex, often involving multi-omics datasets, large-scale cohorts, interdisciplinary approaches and compressed project timelines. All these can promote broader collaboration and increase the prevalence of shared first-author recognition.

\subsubsection*{Byline positions of equal-contributing authors}

To examine where equal-contributing authors appear in the byline, we used the contributor identifiers encoded in the metadata together with the total number of authors per article. For each article with at least one equal-contribution designation, we parsed the \texttt{equal-contribution} author indices (order in the author list) and mapped each identifier $k$ to its integer position $p_{i}$ in the byline, with $1 \leq p_{i} \leq \texttt{num\_authors}$. 

To make positions comparable across articles with different team sizes, we normalized each position to the unit interval,
$$
r_{i} \;=\; 
\begin{cases}
0.5, & \text{if } \texttt{num\_authors}=1,\\[4pt]
\displaystyle \frac{p_{i}-1}{\texttt{num\_authors}-1}, & \text{otherwise},
\end{cases}
$$
so that $r_{i}=0$ corresponds to the first author, $r_{i}=1$ to the last author, and $0<r_{i}<1$ to middle authors. We then pooled all normalized positions $\{r_{i}\}$ over the corpus and summarized their distribution using histograms to characterize the prevalence of co-first, co-middle, and co-last equal-contribution designations.

\begin{figure}[htbp]
  \centering
  \includegraphics[width=0.95\linewidth]{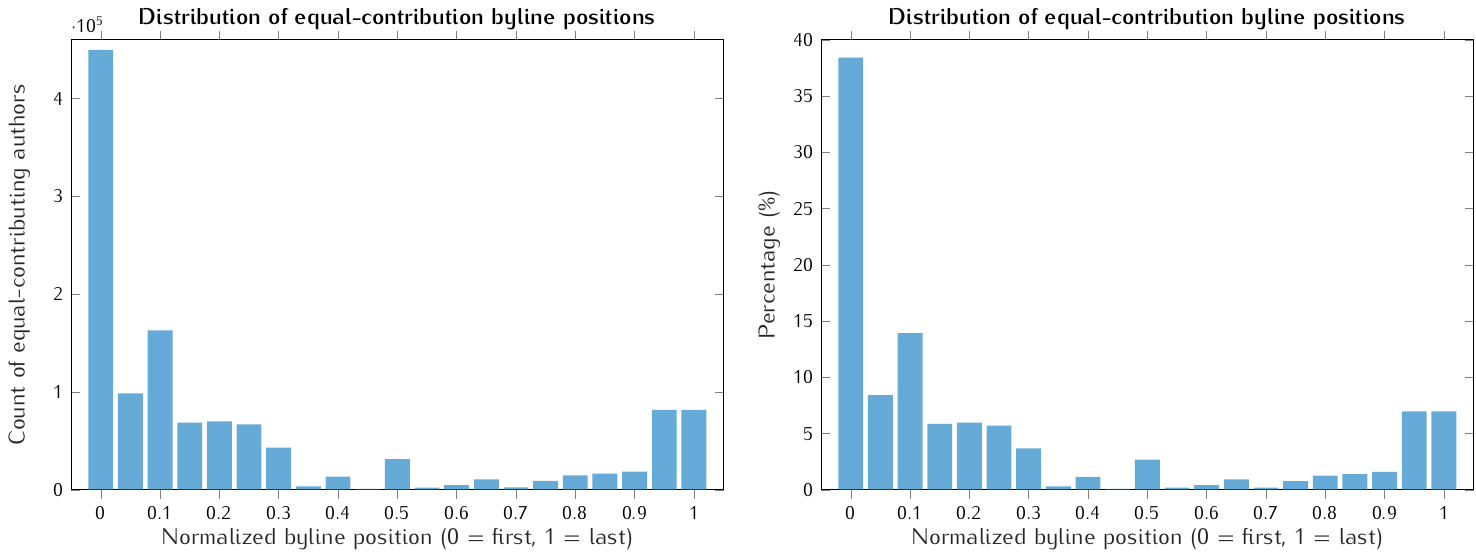}
  \caption{Distribution of byline positions among equal-contributing authors. 
  The normalized position is defined as $(p-1)/(\texttt{num\_authors}-1)$, 
  so that 0 corresponds to the first author and 1 to the last author. 
  \emph{Left}: absolute counts of equal-contributing authors by normalized position. 
  \emph{Right}: the same distribution expressed as percentages.}
  \label{fig:eqc_byline_positions}
\end{figure}

Figure~\ref{fig:eqc_byline_positions} shows that equal-contribution designations are concentrated at the start of the byline: most tagged authors occupy positions very close to the first author ($\sim 38\%$), with a much smaller mass in middle positions and a secondary concentration near the last author. This pattern indicates that equal-contribution labels are used predominantly for co-first authors, with relatively fewer cases of co-middle or co-senior equal-contribution authorship.

\subsection{Journal-level Prevalence Patterns}

We analyze adoption patterns at the journal level rather than by research field or institution. Journals are a natural and logical unit for this phenomenon because editorial policies, author instructions, and local review practices could directly reflect how equal-contribution designations are used. In addition, journal titles and identifiers are relatively stable and consistently indexed in PubMed and PMC, whereas robust field assignments would require additional modeling (e.g., topic classification or MeSH(Medical Subject Headings)-based clustering) and can be sensitive to boundary choices. Institution-level analyses, while of interest, are complicated by the heterogeneity and temporal instability of affiliation strings and by the fact that only a small subset of institutions have enough tagged articles in this corpus to support stable and reliable estimation. For these reasons, we treat journals as the primary unit for describing variation in equal-contribution authorship.

In the first results section, we examined which journals publish the most \texttt{equal-contribution} articles in absolute numbers. This provides an indication of scale but does not show how individual journals have changed over time. Some journals have used this designation for many years, while others appear to have adopted it more recently, with rapid increases thereafter.

\begin{figure}[!htb]
    \centering
    \includegraphics[width=0.5\linewidth]{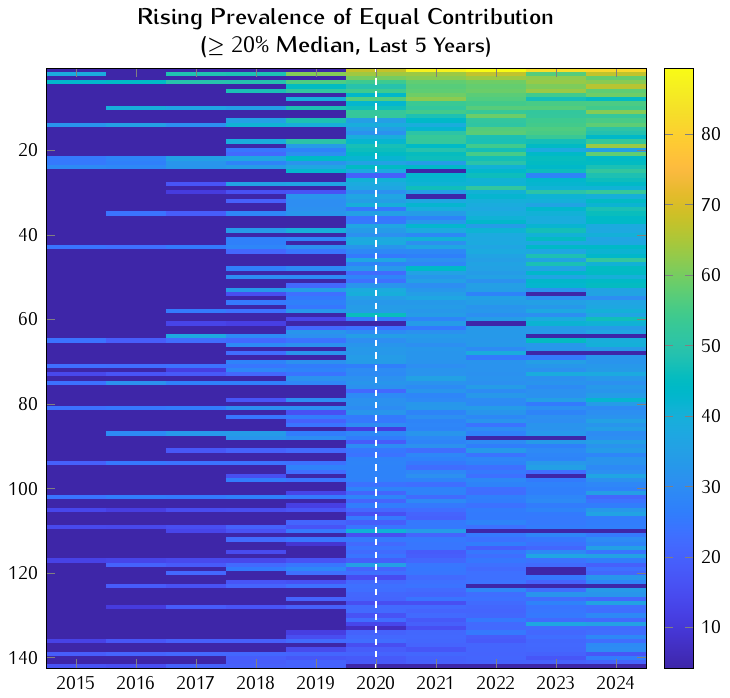}
    \caption{Changes in \texttt{equal-contribution} authorship over time at the journal level.
Heatmap of 143 journals with a median proportion of \texttt{equal-contribution} articles $\geq 20\%$ in the last five years (2020-2024); colors indicate yearly percentages. These values are descriptive only and should not be interpreted as evaluative of editorial policies or journal quality.}
    \label{fig:journal_longitudinal_patterns}
\end{figure}
\FloatBarrier

Figure~\ref{fig:journal_longitudinal_patterns} shows a heatmap for 143 journals in which  the median share of {equal-contribution} articles over the last five years is at least 20\%. Two main patterns can be observed. Some journals have maintained high levels throughout the period, while others began at relatively low levels and then rose sharply, mostly after 2018. These increases may reflect changes in journal policy, shifts in editorial practices, or the influence of large-scale collaborative articles.

Over the last five years, 33 journals exhibited median equal-contribution prevalence of $\geq 40\%$. Among these, the highest individual journal reached 83.2\%, with several others consistently above 60\% and a substantial group in the 40-55\% range. Many of these outlets are concentrated in oncology, cell biology, molecular medicine, and translational research, areas characterized by multi-center studies, diverse expertise, and team-based workflows.

The presence of journals with $\geq 50\%$ prevalence indicates that, in certain research communities, equal-contribution authorship has shifted from an occasional designation to a routine convention. In contrast, large multidisciplinary journals show elevated ratios around 40\%, but with broader variation reflecting diverse disciplinary norms. These findings suggest that both domain-specific practices and editorial conventions influence the adoption of equal-contribution credit.

These patterns are consistent with broader trends toward larger, more interdisciplinary research teams, as discussed in Section~\ref{sec:discussion}.

\begin{figure}[!htb]
    \centering
    \includegraphics[width=.9\linewidth]{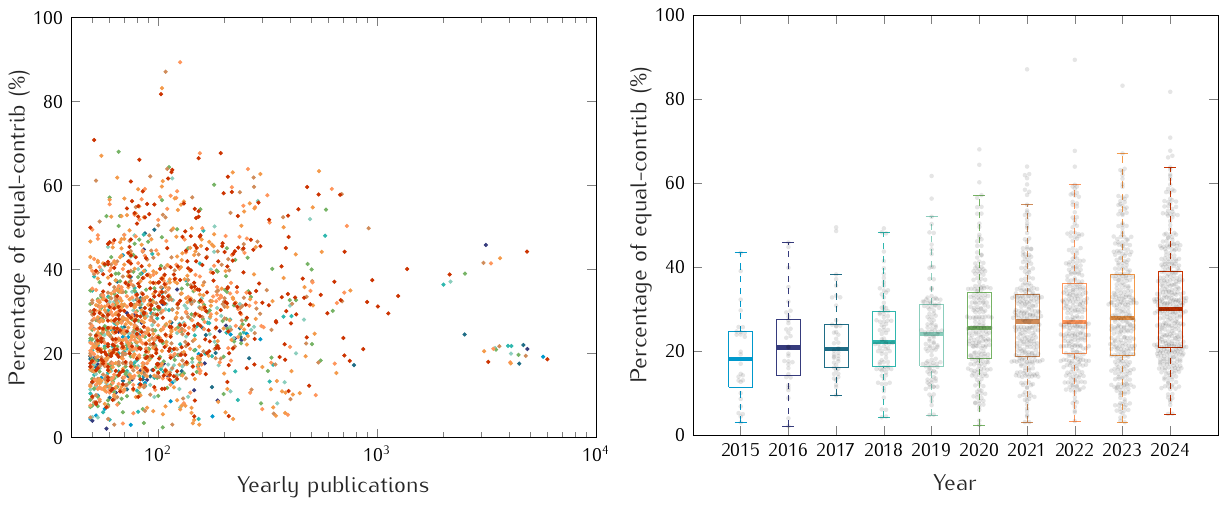}
    \caption{Relationship between annual journal output and the proportion of \texttt{equal-contribution} authorship (\emph{Left}), and yearly distribution of \texttt{equal-contribution} ratios across all journals (\emph{Right}). The scatter plot shows individual journal-year points, color-coded by year; the boxplots summarize the spread and median ratio for 2015-2024.}
    \label{fig:journalyear_pub_counts_Top_boxplot}
\end{figure}
\FloatBarrier

Across all journals, the yearly boxplots (Figure~\ref{fig:journalyear_pub_counts_Top_boxplot}) showed an upward shift in the proportion of papers with equal-contribution authorship. The median ratio was increased from about 19.2\% in 2015 to 30.1\% in 2024. The widening interquartile range indicates that not only is the \emph{typical} adopting journal more likely to use this practice, but the variability in adoption rates between journals is also growing. The scatter plot on the left shows no simple relationship between a journal's annual output and its equal-contribution ratio: high ratios occur in both low-volume and high-volume journals, although extreme values tend to appear in lower-output titles.

Overall, these results suggest that the rise in equal-contribution authorship is a broad trend spanning journals of different sizes, with certain fields and editorial policies driving ratios well above the global median.

\subsection{Geographic Pattern}

To examine the geographic distribution of participation in publications that include equal-contribution designations, we attribute each such article to countries based on the full set of author affiliations. Because team sizes vary substantially, simple counting would over-represent large multi-author papers. We therefore apply a fractional counting scheme in which each article contributes a total weight of one, distributed across countries according to their representation in the author affiliation list. This approach measures country participation in equal-contribution articles at the publication level, rather than the nationality of equal-contributing authors specifically.

\subsubsection*{Computation of country ratios}

For each article with at least one equal-contribution author, all unique countries present in the author affiliation list were extracted (using the ISO 3166-1 mapping procedure described in the Data section). To avoid over-weighting articles with many authors from the same country, we applied a fractional counting approach. Each country was assigned a weight $w_{i,c}$ for article $i$:
\[
w_{i,c} \;=\; \frac{n_{i,c}}{\sum_{c'} n_{i,c'}} ,
\]
where $n_{i,c}$ is the number of occurrences of country $c$ in article $i$'s affiliation list (after parsing and ISO 3166-1 mapping). By construction, $w_{i,c} \geq 0$, $w_{i,c} = 0$ if $n_{i,c} = 0$, and $\sum_c w_{i,c} = 1$ for each article $i$.

We then aggregated these per-article weights over all articles to obtain country totals and proportions at the corpus level.

\subsubsection*{General geographic pattern}

\begin{figure}[!htb]
    \centering
    \includegraphics[width=\linewidth]{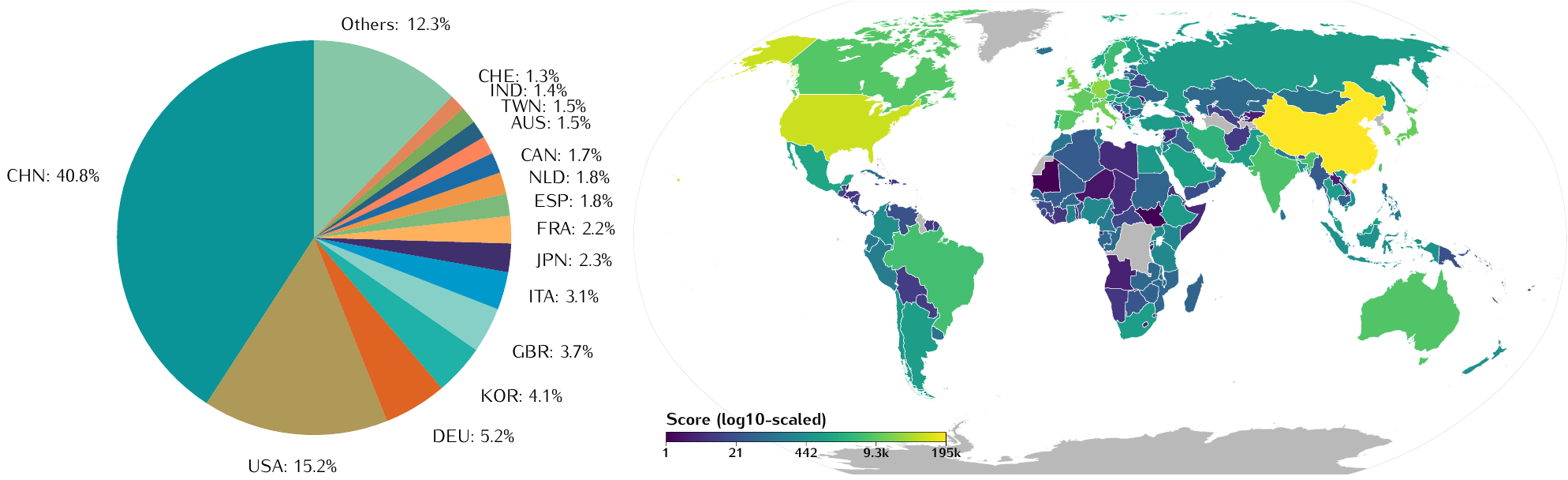}
    \caption{Geographic distribution of \texttt{equal-contribution} authorship.\\
    \emph{Left}: Pie chart showing the fractional share of the top 15 contributing countries or regions plus ``Others''.
    \emph{Right}: World map with log-scaled color mapping of article counts per country or region.}
    \label{fig:geopattern}
\end{figure}

Figure~\ref{fig:geopattern} shows the geographic distribution of equal-contribution authorship, the World Map is obtained from \cite{NaturalEarth2025150m}. Results indicate a high concentration, with China accounting for 40.8\% of all fractionalized contributions, followed by the United States (15.2\%) and Germany (5.2\%). South Korea (4.1\%), the United Kingdom (3.7\%), and Italy (3.1\%) also feature prominently. The remaining top-15 countries or regions such as Japan, France, Spain, the Netherlands, Canada, Australia, Taiwan, India, and Switzerland—each contribute between approximately 1.3\% and 2.3\%. Collectively, the ``Others'' category, which includes all remaining countries, accounts for 12.3\% of the total.

\FloatBarrier

This distribution could reflect both the global research landscape and specific disciplinary trends. The particularly high proportion from China may be associated with the scale of biomedical research output in the country, the frequent use of multi-institution collaborations, and cultural or institutional norms favoring shared first authorship. The United States' high share mirrors its overall research volume but is proportionally smaller than China's, suggesting differences in authorship practices. The comparatively high shares for South Korea and Germany may indicate active participation in multi-center studies in high-ratio domains like oncology and molecular biology. The concentration in a small set of countries also suggests uneven adoption of equal-contribution designations globally, potentially reflecting disparities in funding, collaborative infrastructure, and editorial policies.

\begin{figure}[!h]
    \centering
    \includegraphics[width=\linewidth]{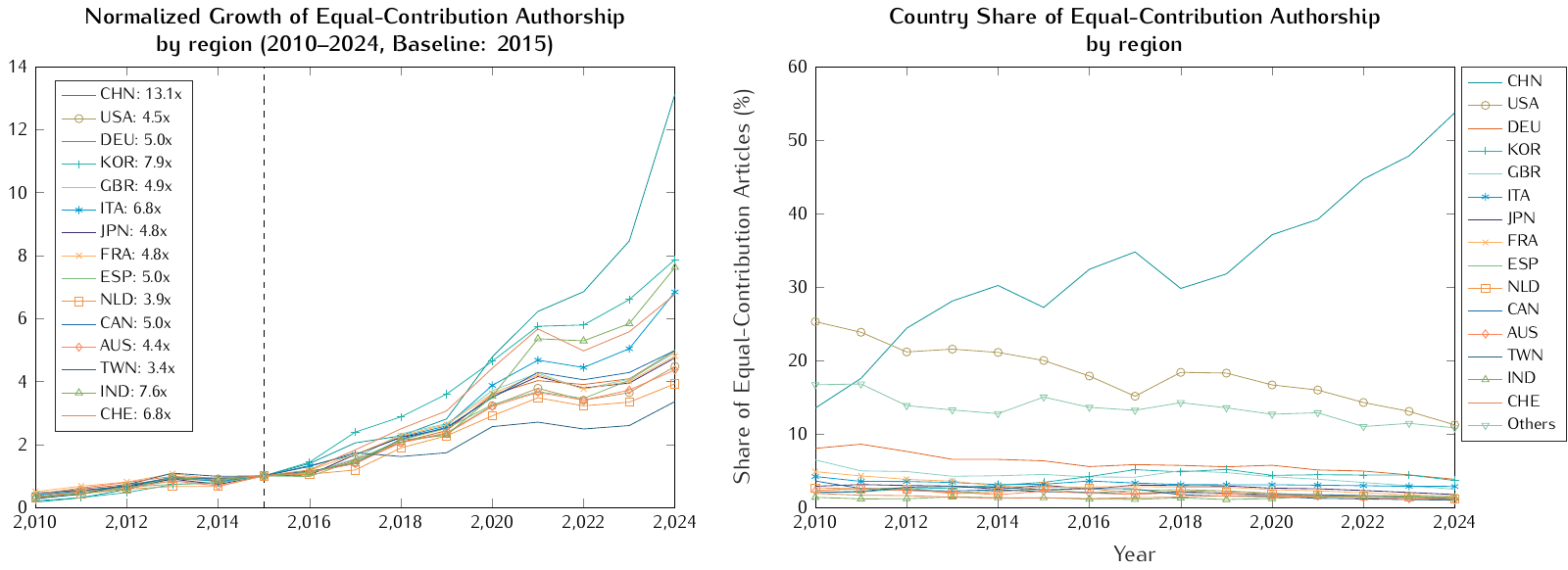}
    \caption{\emph{Left}: Relative increase in publications with \texttt{equal-contribution} authorship across the top 15 contributing countries or regions. All values are normalized to each region's 2015 level. \emph{Right}: Country Share of Equal-Contribution Authorship by region.}
    \label{fig:growth_rate_countries}
\end{figure}

\subsubsection*{Growth pattern over time}

When we normalize each country or region's yearly equal-contribution counts to its own 2015 baseline, the growth patterns become much clearer. 
Although Figure~\ref{fig:growth_rate_countries} (Left) displays data from 2010-2024, we use 2015 as the normalization point to focus on the most recent decade, where data coverage is more complete and consistent across countries. Values from 2010-2014 are retained in the plot for context, but these earlier years generally have lower coverage and should be interpreted with caution.

China shows the steepest increase, with a $13.1\times$ rise by 2024, nearly double that of the next fastest-growing countries, such as South Korea ($7.9\times$), India ($7.6\times$), and Italy and Switzerland (both $6.8\times$). Many European countries, the United States, and Japan cluster in the $4-5\times$ range, while Taiwan ($3.4\times$) and the Netherlands ($3.9\times$) are at the lower end, yet still exhibit substantial growth. 

In terms of geographic distribution of equal-contribution authorship over time, as shown in Figure~\ref{fig:growth_rate_countries} (Right), China's share rises sharply, from about 14\% in 2010 to over 50\% by 2024, making it the dominant contributor by the end of the period. In contrast, the United States shows a steady decreasing in relative share, while Germany and several European countries remain comparatively stable or slightly decreasing. Countries such as South Korea and India show moderate growth but remain below China's level. The decline of the “Others” category suggests increasing concentration among the top contributors. Overall, the pattern reflects a substantial redistribution of global participation in shared first authorship.

These results indicate that the rise in shared first-authorship attribution is not confined to a few regions; even the slowest-growing group has more than tripled its baseline levels over the last decade, suggesting widespread and accelerating adoption of {equal-contribution} practices across the global research landscape.

\section{Discussion and Conclusion}
\label{sec:discussion}

This study used approximately 480\,000 publications tagged with equal-contribution authorship to characterize how this authorship convention has evolved over time, across journals, and across countries. At the global level, we observed a pronounced increase in equal-contribution designations since the mid-2010s, with a particularly steep rise after 2017. The growth appears both in absolute counts and in the proportion of tagged articles at the journal level, suggesting that equal-contribution authorship is becoming a routine feature of bylines in many parts of the biomedical literature rather than an exceptional designation.

The journal-level results indicate substantial heterogeneity. A subset of journals shows consistently high prevalence of equal-contribution authorship over the whole observation period, while others exhibit an increase only in recent years. In the most extreme cases, more than half of all articles in a journal carry an equal-contribution tag. Such patterns would be consistent with the idea that journal scope and editorial policy shape local norms for when equal-contribution is used. They also imply that readers and evaluators cannot assume a single, uniform meaning for these designations across journals: in some journals, equal-contribution appears as a relatively rare signal; in others, it is close to the default.

The geographic analysis shows a similar pattern. China now accounts for the largest share (40.8\%) of equal-contribution articles in the corpus, followed by the  United States (15.2\%) and Germany (5.2\%), with several other countries contributing smaller but non-negligible shares. When we normalize each country or region to its own 2015 baseline, nearly all of them show at least a threefold increase by 2024, with particularly strong relative growth in China, South Korea, and India. This indicates that the spread of equal-contribution authorship is not confined to a small subset of countries, although the levels and timing of adoption differ. In this sense, the rise of equal-contribution authorship is both a global and a geographically uneven phenomenon.

Equal-contribution authorship has practical consequences because it interacts directly with the way credit and responsibility are inferred from author order. In principle, co-first or jointly credited authorship allows committees, funders, and hiring panels to recognize genuinely shared leadership in projects that would be difficult to summarize with a strictly linear byline. At the same time, qualitative and editorial-based studies have reported that the interpretation of joint-first and equal-contribution labels is not uniform across institutions and evaluation contexts. Some authors perceive these labels as necessary to acknowledge collaborative work fairly, while others report uncertainty about how such designations are counted in CVs and promotion processes. Against this background, the high prevalence we document in some journals and the rapid growth in several countries suggest that equal-contribution has become a widely used signal whose practical meaning may depend strongly on local norms and policies rather than on a single commonly shared understanding.

More broadly, the results can be seen as part of a longer development in scientific publishing, where authorship practices change in response to larger teams, evaluation incentives, and editorial rules. The same publication system that has produced larger and more specialized teams, as well as a growing group of middle authors, has also introduced new ways to describe who did what, such as contributorship taxonomies and explicit equal-contribution labels. The patterns we report do not tell us which individual uses of these labels are appropriate or inappropriate. They do show, however, that one specific signal, equal-contribution authorship, has grown quickly and unevenly across the biomedical literature. This raises a practical question for journals and evaluators: how to keep author order and contribution labels informative when they are widely used and applied in different ways across journals.

From a policy perspective, these results suggest some possible implications, rather than clear recommendations. First, the frequent use of equal-contribution labels indicates that journals may need clear and detailed policies on when such labels are appropriate and how they should be connected to contributorship statements. Second, if co-first and equal-contribution authorship becomes common in some venues, it may no longer function as a rare signal of exceptional joint leadership. In that case, evaluators may need to place more weight on structured contribution information (for example, CRediT roles) and relatively less on small differences in author order. Third, regular monitoring of authorship patterns at the journal or publisher level, using metadata of the type analyzed here, could help detect unintended shifts in local practices and provide an empirical basis for revising authorship policies when needed.

\subsection*{Limitations}

This study has several limitations that should be considered when interpreting the results. First, the analysis relies on structured metadata from PubMed and PubMed Central, in particular on JATS elements and attributes such as \texttt{equal-contrib="yes"}. Equal-contribution statements that appear only in unstructured text (for example, in PDF footnotes or narrative acknowledgments) are not captured, so the corpus used here represents a lower bound on the true prevalence of equal-contribution authorship. Coverage of full JATS metadata also varies across journals and over time, which may affect the apparent timing of adoption for some venues.

Second, the metadata themselves are not error-free. Inconsistent tagging by publishers, occasional parsing errors and heterogeneous use of journal and affiliation fields can introduce noise. We tried to reduce these issues through conservative cleaning and matching rules, but some misclassification is likely to happen. Affiliation-based geocoding adds another source of uncertainty: affiliation strings are heterogeneous, can change over time for the same institution, and do not always map cleanly to a single country. The country-level results should therefore be interpreted as approximate, fractional indicators of participation in equal-contribution articles rather than precise counts.

Third, the analysis is restricted to articles that contain explicit equal-contribution tags and does not normalize by the full publication output of each journal or country over the same period. For journals and countries with rapidly changing overall output, the measures reported here describe the structure of equal-contribution authorship within the tagged subset, not comprehensive authorship rates across all publications. Finally, we focus on variation across journals and countries. A more detailed analysis by research field, institution, or research team would require additional disambiguation and classification steps (for example, topic modeling or affiliation reconciliation), which are beyond the scope of the present study. 

Despite these limitations, the size of the corpus and the use of machine-readable metadata provide a useful basis for identifying large-scale patterns in the sharp rise of equal-contribution authorship. The results should be read as a descriptive baseline that future work can refine and extend using more detailed domain-specific, institutional, or qualitative data.

\section*{Competing Interests}

The author has no conflicting interests.

\section*{Funding Information}

This study received no specific grant. 

\bibliographystyle{unsrturl}
\bibliography{bib}

@ARTICLE{Hanson2024strain,
  author = {Hanson, Mark A. and Barreiro, Pablo Gómez and Crosetto, Paolo and Brockington, Dan},
  title = {The strain on scientific publishing},
  journal = {Quantitative Science Studies},
  year = {2024},
  volume = {5},
  pages = {823-843},
  number = {4},
  month = {11},
  eprint = {https://direct.mit.edu/qss/article-pdf/5/4/823/2478590/qss\_a\_00327.pdf},
  doi = {10.1162/qss_a_00327},
  issn = {2641-3337},
  url = {10.1162/qss\_a\_00327}
}

@ARTICLE{Jakab2024How,
  author = {Jakab, Martin and Kittl, Eva and Kiesslich, Tobias},
  title = {How many authors are (too) many? A retrospective, descriptive analysis of authorship in biomedical publications},
  journal = {Scientometrics},
  year = {2024},
  volume = {129},
  pages = {1299--1328},
  number = {3},
  doi = {10.1007/s11192-024-04928-1},
  issn = {1588-2861},
}

@ARTICLE{Wuchty2007Increasing,
  author = {Stefan Wuchty and Benjamin F. Jones and Brian Uzzi},
  title = {The Increasing Dominance of Teams in Production of Knowledge},
  journal = {Science},
  year = {2007},
  volume = {316},
  pages = {1036-1039},
  number = {5827},
  eprint = {https://www.science.org/doi/pdf/10.1126/science.1136099},
  doi = {10.1126/science.1136099},
  url = {https://www.science.org/doi/abs/10.1126/science.1136099}
}

@ARTICLE{Akhabue2010Equal,
  author = {Ehimare Akhabue and Ebbing Lautenbach},
  title = {“Equal” Contributions and Credit: An Emerging Trend in the Characterization of Authorship},
  journal = {Annals of Epidemiology},
  year = {2010},
  volume = {20},
  pages = {868-871},
  number = {11},
  doi = {10.1016/j.annepidem.2010.08.004},
  issn = {1047-2797},
  url = {https://www.sciencedirect.com/science/article/pii/S104727971000308X}
}

@ARTICLE{Huang2016co,
  author = {Huang, Mu-hsuan and Hsieh, Hsiao-Ting and Lin, Chi-Shiou},
  title = {The co-first and co-corresponding author phenomenon in the pharmacy and anesthesia journals},
  journal = {Proceedings of the Association for Information Science and Technology},
  year = {2016},
  volume = {53},
  pages = {1-4},
  number = {1},
  doi = {10.1002/pra2.2016.14505301138},
  url = {https://asistdl.onlinelibrary.wiley.com/doi/abs/10.1002/pra2.2016.14505301138}
}

@ARTICLE{Tian2024Understanding,
  author = {Tian, Wencan and Cai, Ruonan and Fang, Zhichao and Geng, Yu and Wang, Xianwen and Hu, Zhigang},
  title = {Understanding co-corresponding authorship: A bibliometric analysis and detailed overview},
  journal = {Journal of the Association for Information Science and Technology},
  year = {2024},
  volume = {75},
  pages = {3-23},
  number = {1},
  doi = {10.1002/asi.24836},
  url = {https://asistdl.onlinelibrary.wiley.com/doi/abs/10.1002/asi.24836}
}

@TECHREPORT{NISOCRTWG2022ANSINISO,
  author = {{NISO CRediT Working Group}},
  title = {{ANSI/NISO} Z39.104-2022, {CRediT}, contributor roles taxonomy},
  institution = {NISO},
  year = {2022},
  address = {Baltimore, MD}
}

@misc{NaturalEarth2025150m,
  author       = {{Natural Earth}},
  title        = {{1:50m Cultural Vectors, v5.1.1}},
  howpublished = {Online},
  year         = {2025},
  note         = {Available from: \url{https://www.naturalearthdata.com/downloads/50m-cultural-vectors/}},
}

@misc{BNLM2025PMC,
  author       = {{National Library of Medicine}},
  title        = {{PMC Open Access Subset}},
  howpublished = {Online},
  year         = {2025},
  note         = {Bethesda (MD): National Library of Medicine; Available from: \url{https://pmc.ncbi.nlm.nih.gov/tools/openftlist/}},
}

@misc{ICMJE2025Defining,
  author       = {International Committee of Medical Journal Editors},
  title        = {Defining the Role of Authors and Contributors},
  howpublished = {Online},
  year         = {2025},
  note         = {Available from: \url{https://www.icmje.org/recommendations/browse/roles-and-responsibilities/defining-the-role-of-authors-and-contributors.html}},
}

@misc{NaturePortfolio2025Authorship,
  author       = {Nature Portfolio},
  title        = {{Authorship Policy}},
  howpublished = {Online},
  year         = {2025},
  note         = {Available from: \url{https://www.nature.com/nature-portfolio/editorial-policies/authorship}},
}

@misc{BMJ2025Authorship,
  author       = {{BMJ}},
  title        = {{Authorship \& Contributorship}},
  howpublished = {Online},
  year         = {2025},
  note         = {Available from: \url{https://www.bmj.com/about-bmj/resources-authors/article-submission/authorship-contributorship}},
}

@misc{PLOS2025Authorship,
  author       = {{PLOS}},
  title        = {Authorship Policy},
  howpublished = {Online},
  year         = {2025},
  note         = {Available from: \url{https://journals.plos.org/plosone/s/authorship}},
}

@misc{Central2025PubMed,
  author       = {{PubMed Central}},
  title        = {{PubMed Central Tagging Guidelines}},
  howpublished = {Online},
  year         = {2025},
  note         = {Available from: \url{https://www.ncbi.nlm.nih.gov/pmc/pmcdoc/tagging-guidelines/article/tags.html}},
}

@MISC{ICMJE2025Recommendations,
  author = {{International Committee of Medical Journal Editors}},
  title = {{Recommendations for the Conduct, Reporting, Editing, and Publication of Scholarly Work in Medical Journals}},
  year = {2025},
  howpublished = {Online},
  month = {April},
  note = {Updated April 2025. Available from: \url{https://www.icmje.org/recommendations/}},
}

@ARTICLE{Beck2011NISO,
  author = {Beck, Jeffrey},
  title = {{NISO Z39. 96 The Journal Article Tag Suite (JATS): What Happened to the NLM DTDs?}},
  journal = {The journal of electronic publishing: JEP},
  year = {2011},
  volume = {14},
  pages = {106},
  number = {1},
}

@ARTICLE{Lariviere2015Team,
  author = {Larivière, Vincent and Gingras, Yves and Sugimoto, Cassidy R. and Tsou, Andrew},
  title = {Team size matters: Collaboration and scientific impact since 1900},
  journal = {Journal of the Association for Information Science and Technology},
  year = {2015},
  volume = {66},
  pages = {1323-1332},
  number = {7},
  eprint = {https://asistdl.onlinelibrary.wiley.com/doi/pdf/10.1002/asi.23266},
  doi = {10.1002/asi.23266},
  url = {https://asistdl.onlinelibrary.wiley.com/doi/abs/10.1002/asi.23266}
}

@ARTICLE{Fortunato2018Science,
  author = {Santo Fortunato and Carl T. Bergstrom and Katy Börner and James A. Evans and Dirk Helbing and Staša Milojević and Alexander M. Petersen and Filippo Radicchi and Roberta Sinatra and Brian Uzzi and Alessandro Vespignani and Ludo Waltman and Dashun Wang and Albert-László Barabási },
  title = {Science of science},
  journal = {Science},
  year = {2018},
  volume = {359},
  pages = {eaao0185},
  number = {6379},
  eprint = {https://www.science.org/doi/pdf/10.1126/science.aao0185},
  doi = {10.1126/science.aao0185},
  url = {https://www.science.org/doi/abs/10.1126/science.aao0185}
}

@ARTICLE{Cronin2001Hyperauthorship,
  author = {Cronin, Blaise},
  title = {Hyperauthorship: A postmodern perversion or evidence of a structural shift in scholarly communication practices?},
  journal = {Journal of the American Society for Information Science and Technology},
  year = {2001},
  volume = {52},
  pages = {558-569},
  number = {7},
  eprint = {https://onlinelibrary.wiley.com/doi/pdf/10.1002/asi.1097},
  doi = {10.1002/asi.1097},
  url = {https://onlinelibrary.wiley.com/doi/abs/10.1002/asi.1097}
}

@ARTICLE{Fox2016Citations,
  author = {Fox, Charles W. and Paine, C. E. Timothy and Sauterey, Boris},
  title = {Citations increase with manuscript length, author number, and references cited in ecology journals},
  journal = {Ecology and Evolution},
  year = {2016},
  volume = {6},
  pages = {7717-7726},
  number = {21},
  eprint = {https://onlinelibrary.wiley.com/doi/pdf/10.1002/ece3.2505},
  doi = {10.1002/ece3.2505},
  url = {https://onlinelibrary.wiley.com/doi/abs/10.1002/ece3.2505}
}

@ARTICLE{Mongeon2017rise,
  author = {Mongeon, Philippe AND Smith, Elise AND Joyal, Bruno AND Larivière, Vincent},
  title = {The rise of the middle author: Investigating collaboration and division of labor in biomedical research using partial alphabetical authorship},
  journal = {PLOS ONE},
  year = {2017},
  volume = {12},
  pages = {1-14},
  number = {9},
  month = {09},
  doi = {10.1371/journal.pone.0184601},
  publisher = {Public Library of Science},
  url = {10.1371/journal.pone.0184601}
}

@ARTICLE{Rennie1994Authorship,
  author = {Rennie, Drummond and Flanagin, Annette},
  title = {Authorship! Authorship!: Guests, Ghosts, Grafters, and the Two-Sided Coin},
  journal = {JAMA},
  year = {1994},
  volume = {271},
  pages = {469-471},
  number = {6},
  month = {02},
  doi = {10.1001/jama.1994.03510300075043},
  eprint = {https://jamanetwork.com/journals/jama/articlepdf/364559/jama_271_6_043.pdf},
  issn = {0098-7484},
  url = {10.1001/jama.1994.03510300075043}
}

@ARTICLE{Flanagin1998Prevalence,
  author = {Flanagin, Annette and Carey, Lisa A. and Fontanarosa, Phil B. and Phillips, Stephanie G. and Pace, Brian P. and Lundberg, George D. and Rennie, Drummond},
  title = {Prevalence of Articles With Honorary Authors and Ghost Authors in Peer-Reviewed
Medical Journals},
  journal = {JAMA},
  year = {1998},
  volume = {280},
  pages = {222-224},
  number = {3},
  month = {07},
  eprint = {https://jamanetwork.com/journals/jama/articlepdf/187772/jpv80004.pdf},
  doi = {10.1001/jama.280.3.222},
  issn = {0098-7484},
  url = {10.1001/jama.280.3.222}
}

@ARTICLE{Kornhaber2015Ongoing,
  author = {Kornhaber, Rachel and McLean, Loyola and Baber, Rodney},
  title = {Ongoing ethical issues concerning authorship in\&amp;nbsp;biomedical journals: an integrative review},
  journal = {International Journal of Nanomedicine},
  year = {2015},
  pages = {4837},
  month = jul,
  doi = {10.2147/ijn.s87585},
  issn = {1178-2013},
  publisher = {Informa UK Limited},
  url = {http://dx.doi.org/10.2147/IJN.S87585}
}

@ARTICLE{Owens2024Equality,
  author={Owens, R{\'o}is{\'\i}n M and Simmonds, Liz and Malliaras, George},
  title = {Equality in publishing: Are joint authors truly equal?},
  journal = {Science Advances},
  year = {2024},
  volume = {10},
  number = {30},
  month = jul,
  doi = {10.1126/sciadv.adq9382},
  issn = {2375-2548},
  publisher = {American Association for the Advancement of Science (AAAS)},
  url = {http://dx.doi.org/10.1126/sciadv.adq9382}
}

@ARTICLE{Efron2024Joint,
  author = {Efron, Nathan},
  title = {Joint first authorship},
  journal = {Clinical and Experimental Optometry},
  year = {2024},
  volume = {107},
  pages = {243–244},
  number = {3},
  month = apr,
  doi = {10.1080/08164622.2022.2146486},
  issn = {1444-0938},
  publisher = {Informa UK Limited},
  url = {http://dx.doi.org/10.1080/08164622.2022.2146486}
}

@ARTICLE{Biagioli2012Recycling,
  author = {Biagioli, Mario},
  title = {Recycling Texts or Stealing Time?: Plagiarism, Authorship, and Credit in Science},
  journal = {International Journal of Cultural Property},
  year = {2012},
  volume = {19},
  pages = {453–476},
  number = {3},
  month = aug,
  doi = {10.1017/s0940739112000276},
  issn = {1465-7317},
  publisher = {Cambridge University Press (CUP)},
  url = {http://dx.doi.org/10.1017/S0940739112000276}
}

@ARTICLE{Khalifa2022Losing,
  author = {Khalifa, A.A.},
  title = {Losing young researchers in the authorship battle, under-reported casualties},
  journal = {Ethics, Medicine and Public Health},
  year = {2022},
  volume = {20},
  pages = {100735},
  month = feb,
  doi = {10.1016/j.jemep.2021.100735},
  issn = {2352-5525},
  publisher = {Elsevier BV},
  url = {http://dx.doi.org/10.1016/j.jemep.2021.100735}
}

@ARTICLE{Kwok2005White,
  author = {Kwok, L S},
  title = {The White Bull effect: abusive coauthorship and publication parasitism},
  journal = {Journal of Medical Ethics},
  year = {2005},
  volume = {31},
  pages = {554–556},
  number = {9},
  month = aug,
  doi = {10.1136/jme.2004.010553},
  issn = {1473-4257},
  publisher = {BMJ},
  url = {http://dx.doi.org/10.1136/jme.2004.010553}
}

@ARTICLE{Macfarlane2015ethics,
  author = {Macfarlane, Bruce},
  title = {The ethics of multiple authorship: power, performativity and the gift economy},
  journal = {Studies in Higher Education},
  year = {2015},
  volume = {42},
  pages = {1194–1210},
  number = {7},
  month = oct,
  doi = {10.1080/03075079.2015.1085009},
  issn = {1470-174X},
  publisher = {Informa UK Limited},
  url = {http://dx.doi.org/10.1080/03075079.2015.1085009}
}

@article{PardalRefoyo2025Gift,
  title = {Gift,  guest \& ghost authorship in biomedical publications: definitions,  prevalence,  impacts,  detection and prevention. Scoping review},
  url = {http://dx.doi.org/10.21203/rs.3.rs-6577822/v1},
  doi = {10.21203/rs.3.rs-6577822/v1},
  publisher = {Springer Science and Business Media LLC},
  author = {Pardal-Refoyo,  José Luis and Pardal-Peláez,  Beatriz},
  year = {2025},
  month = may 
}

@ARTICLE{Hosseini2020qualitative,
  author = {Hosseini, Mohammad and Bruton, Samuel V.},
  title = {A qualitative study of Equal Co-First Authorship},
  journal = {Accountability in Research},
  year = {2020},
  volume = {27},
  pages = {496–520},
  number = {8},
  month = jun,
  doi = {10.1080/08989621.2020.1776122},
  issn = {1545-5815},
  publisher = {Informa UK Limited},
  url = {http://dx.doi.org/10.1080/08989621.2020.1776122}
}

@ARTICLE{Biagioli2022Ghosts,
  author = {Mario Biagioli},
  title = {Ghosts, brands, and influencers: Emergent trends in scientific authorship},
  journal = {Social Studies of Science},
  year = {2022},
  volume = {52},
  pages = {463-487},
  number = {3},
  note = {PMID: 35491951},
  eprint = {10.1177/03063127221095046},
  doi = {10.1177/03063127221095046},
}

@dataset{xu_2025_16968197,
  author       = {Xu, Binbin},
  title        = {Dataset for ``{Trends in Equal-Contribution Authorship: A Large-Scale Bibliometric Analysis of Biomedical Literature}"},
  month        = aug,
  year         = 2025,
  publisher    = {Zenodo},
  doi          = {10.5281/zenodo.16968197},
  url          = {https://doi.org/10.5281/zenodo.16968197},
}

\end{document}